\documentclass{article}
\usepackage{spconf,amsmath,graphicx,hyperref}
\usepackage{booktabs}            
\usepackage{siunitx}             
\usepackage{multirow}
\usepackage{caption}
\usepackage{enumitem}
\usepackage{stfloats}
\usepackage{amssymb}
\usepackage{makecell}   
\usepackage{subcaption} 

\renewcommand{\thetable}{\Roman{table}}    

\title{TFGA-Net: Temporal-Frequency Graph Attention Network for Brain-Controlled Speaker Extraction}
\name{
  Youhao Si, Yuan Liao, Qiushi Han, Yuhang Yang, Rui Dai and Liya Huang$^\dag$\thanks{$^\dag$ This is the Corresponding Author.}
}
\address{%
    College of Electronic and Optical Engineering \& College of Flexible Electronics (Future Technology), \\
    Nanjing University of Posts and Telecommunications, Nanjing, China
}

\begin{document}
\ninept
\maketitle
\begin{abstract}
The rapid development of auditory attention decoding (AAD) based on electroencephalography (EEG) signals offers the possibility EEG-driven target speaker extraction. However, how to effectively utilize the target-speaker common information between EEG and speech remains an unresolved problem. In this paper, we propose a model for brain-controlled speaker extraction, which utilizes the EEG recorded from the listener to extract the target speech. In order to effectively extract information from EEG signals, we derive multi-scale time--frequency features and further incorporate cortical topological structures that are selectively engaged during the task. Moreover, to effectively exploit the non-Euclidean structure of EEG signals and capture their global features, the graph convolutional networks and self-attention mechanism are used in the EEG encoder. In addition, to make full use of the fused EEG and speech feature and preserve global context and capture speech rhythm and prosody, we introduce MossFormer2 which combines MossFormer and RNN-Free Recurrent as separator.
Experimental results on both the public Cocktail Party and KUL dataset in this paper show that our TFGA-Net model significantly outper-forms the state-of-the-art method in certain objective evaluation metrics. The source code is available at: https://github.com/LaoDa-X/TFGA-NET.

\end{abstract}
\section{Introduction}
\label{sec:Introduction}
Selective auditory attention enables listeners to focus on a single talker in multi-speaker environments such as a cocktail party \cite{Bronkhorst2000CocktailParty}, while actively suppressing competing sources. However, individuals with hearing impairment often struggle with this task. Although modern hearing aids integrate front-end algorithms such as noise reduction and speech enhancement \cite{Srinivasan2019DeepSegregation}, they still cannot infer whom the wearer actually intends to listen to. Consequently, endowing machines with human-level selective listening remains a fundamental challenge.

With the advent of deep learning, speech separation (SS) has made continuous progress, from early deep clustering to Conv-TasNet \cite{Luo2019ConvTasNet}, DPRNN \cite{Luo2020DPRNN}, and more recent SepFormer \cite{Subakan2021SepFormer} and TF-GridNet \cite{Wang2023TFGridNet}. Under the assumption that the number of talkers is known, these systems can decompose a mixture into multiple independent channels. Nevertheless, they must separate all potential speakers, resulting in high computational complexity, and the separated outputs are not aligned with the listener’s attentional focus. Downstream modules such as attention detection are still required to identify the target stream, which further increases consumption.

To reduce redundant computation and concentrate on the listener’s object of attention, speaker extraction (SE) has been proposed. It exploit reference cues, such as enrolled target speech, lip movements \cite{Gu2020MMTSS}, or spatial orientation \cite{Gu2019NeuralSpatialFilter} to directly extracts the target speech from the mixture. Although this strategy performs well when reliable priors are available, the practicality of both acoustic and visual reference cues is limited. When reference cues are missing or inaccurate, the practicality of speaker extraction degrades substantially. This limitation motivates the search for alternative modalities that can more robustly reflect the listener’s true focus of attention. 

Recent studies have demonstrated a strong association between brain activity and the speech being attended \cite{Ceolini2020BISS}. Electroencephalography (EEG), a non-invasive and low-cost technique, allows researchers to decode auditory attention of listeners and identify the target speaker. Early work commonly adopted a “blind separation + auditory attention decoding (AAD) \cite{VanEyndhoven2017EEGAttended}” cascade: EEG was first used to estimate the target speech envelope, which was then compared with each separated source to identify the target talker. However, the overall performance was highly dependent on the accuracy of the auditory attention decoding. Moreover, cascaded approaches are prone to error propagation, limiting system reliability.

In this paper, we introduce TFGA-Net, that approach directly models listeners' attentional focus from the recorded EEG signals to extract the target speech. It consists of four components: speech encoder, EEG encoder, Speaker Extraction module, and speech decoder. The EEG encoder captures multi--scale time--frequency signatures and embeds task-selective cortical topology. The speaker extraction integrates MossFormer \cite{Zhao2023MossFormer} with an RNN-free Recurrent, enabling it to retain global context and capture the rhythm and prosodic characteristics of speech.
By combining local feature modeling with long-range contextual information, this architecture provides a balanced mechanism to enhance target speech while suppressing irrelevant sources. Experiments on the Cocktail Party and KUL datasets show that the TFGA-Net model achieves state-of-the-art performance across multiple evaluation metrics, with improvements of 14.1\% and 15.8\% in terms of Scale-Invariant Signal-to-Distortion Ratio (SI-SDR).

The main contributions of this paper are summarized as follows:
\begin{enumerate}[label=(\arabic*),   
                labelsep=0.6em,      
                leftmargin=*,        
                itemsep=0pt,         
                topsep=0pt,          
                parsep=0pt]          
  \item We introduce a novel EEG encoder, which not only extracts multi-scale time--frequency representations of EEG signals but also integrates cortical topological structures that are selectively recruited during the task.

  \item We introduce a new speaker extraction module, which preserves the global context of fused representations and, at the same time, captures the periodicity and prosodic patterns of speech.

  \item We validate the proposed TFGA-Net model through a series of experiments on the Cocktail Party and KUL datasets, which show significant improvements over the baselines.
\end{enumerate}

\section{METHODS}
\label{sec:METHODS}

\subsection{Problem Formulation}
\label{ssec:Problem Formulation}
Let $x(t)$ be a noisy multi-speaker mixture in the time domain:
\begin{equation}
\label{eq:equation1}
x(t) = s_{\text{target}}(t) + \sum_{i=1}^{I} s_{\text{other},i}(t) \in \mathbb{R}^{T}
\end{equation}

where $s_{\mathrm{target}}(t)$ denotes the speech signal of the user-selected target speaker, $s_{\mathrm{other},i}(t)$ denotes the speech of $I$ interfering speakers, and $T$ denotes the time length of mixture speech segments.
\begin{figure*}[!t]      
  \centering
  \includegraphics[width=\textwidth]{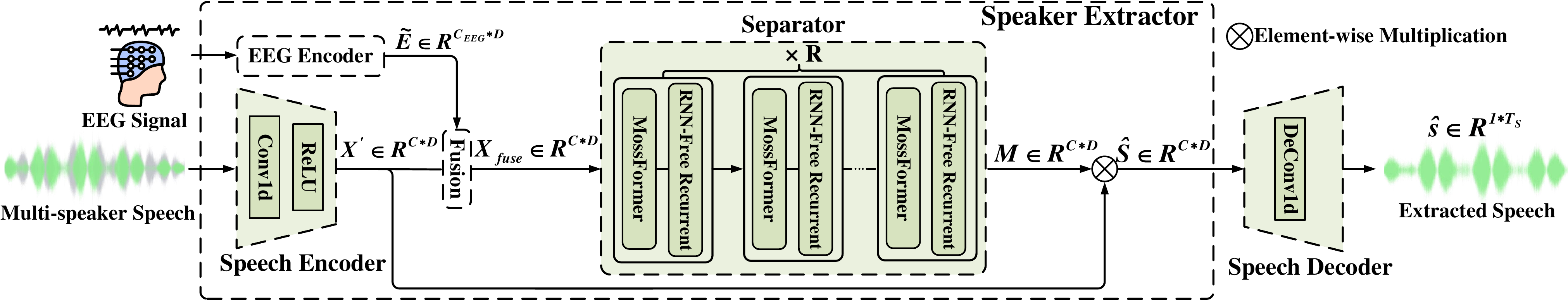}  
  \caption{The overall block diagram of the proposed TFGA-Net model.}
  \label{fig:all}
\end{figure*}
\subsection{Overall Architecture}
\label{ssec:Overall Architecture}
Fig.1 presents the overall structure of TFGA-Net, consisting of the the Speech Encoder, EEG Encoder, the Speaker Extraction Network, and the Speech Decoder.

The following sections provide a detailed explanation of each
component.

\textbf{Speech Encoder.}
The speech encoder consists of a one-dimensional convolutional layer(Conv1D) followed immediately by a ReLU activation function, ensuring that the encoded features remain non-negative. For an input sequence \(X \in \mathbb{R}^{B \times 1 \times T_{S}}\) (where \(B\) is the batch size and \(T_{S}\) is the input length), the encoder applies a kernel size \(K_{1}\) with a stride of \(\tfrac{K_{1}}{2}\), producing an encoded output \(X'\), which can be defined as:
\begin{equation}
\label{eq:equation2}
X' = \operatorname{ReLU}(\operatorname{Conv1D}(X)) \in \mathbb{R}^{B \times C \times D}
\end{equation}
where $N$ denotes the number of filters, and
$S = \tfrac{2\,(T - K_{1})}{K_{1}} + 1$ represents the reduced temporal dimension.

\textbf{EEG Encoder.}
EEG signals encompass rich temporal and frequency characteristics,
and the task-selective responsiveness of distinct cortical regions makes spatial topology equally important. However, current temporal convolution models (TCN \cite{Bai2018TCN} capture short-range patterns but ignore EEG functional connectivity, whereas graph-convolution models (GCN) enhance task-relevant regional activity, yet neglect long-range temporal dependencies.
To address this, we introduce a temporal-frequency graph attention EEG encoding framework. This would allow us to characterize the hierarchical processing of the brain of target speech and provide top-down cues for speaker extraction.

The EEG encoder is designed to learn EEG embedding $\tilde{E}$ from the input EEG signal E that exhibit correlations with the interested speech.

Specifically, for EEG data \(E \in \mathbb{R}^{B \times C \times T_{e}}\)
(where \(B\) is the batch size, \(C\) is the number of electrode
channels, and \(T_{e}\) is the input sequence length),
the signal is sent to two components:
a multi-scale temporal convolution module and a multi-frequency feature extraction module.

In the temporal convolution module, EEG data is processed by one-dimensional convolutional kernels with different receptive fields.
We employ five convolution kernels whose lengths decay exponentially while remaining proportional to the sampling rate:
\(S_{T}^{k} = (1,\,0.5^{k} f_{s})\) , where \(k \in \{1,\dots,5\}\).
Let the output of the \(k^{\text{th}}\) temporal kernel be
\(E_{T}^{k}\in\mathbb{R}^{B\times C\times T\times f_{k}}\),
where \(T\) is the number of temporal kernels, and \(f_{k}\) denotes the
feature length:
\begin{equation}
  E_{T}^{k}
  = \operatorname{ELU}\!\Bigl(
      \operatorname{BN}\bigl(
        \operatorname{Conv1d}\bigl(E, S_{T}^{k}\bigr)
      \bigr)
    \Bigr)
  \tag{3}
\end{equation}
We concatenate the five outputs along the feature dimension and apply a $1{\times}1$ convolution to obtain $E_T$.

In the frequency module, a short-time Fourier transform (STFT) is
applied to each channel. Band-limited power is used to
extract PSD and DE features. The signal is split into the five canonical bands:
$\delta\,(0\text{--}4\,\mathrm{Hz})$,
$\theta\,(4\text{--}8\,\mathrm{Hz})$,
$\alpha\,(8\text{--}12\,\mathrm{Hz})$,
$\beta\,(12\text{--}30\,\mathrm{Hz})$, and
$\gamma\,(30\text{--}50\,\mathrm{Hz})$.
Averaging within each band yields
$E_{p}\!\in\!\mathbb{R}^{C\times D_{F}}$ (PSD) and $E_{D}\!\in\!\mathbb{R}^{C\times D_{F}}$ (DE), where $C$ is the number of channels and $D_{F}=5$. Then, we combine the PSD and DE features to represent the EEG information in the frequency domain, denoted as \(E_{F} \in \mathbb{R}^{C\times D_{F}}.\)

In the next part of this module, we model multi-channel EEG features
using a graph; each electrode in the EEG data is regarded as a node.
To explore the implicit relationships among nodes, we employ
Graph Convolutional Networks (GCNs). The adjacency matrix\(A\)
represents the long--short distance brain network
\(G\) and is symmetric because the graph is undirected.
The initial adjacency matrices for the two views,
\(A_{T}^{\text{initial}}\) and \(A_{F}^{\text{initial}}\),
are set identically to \(A\).
Using this construction, we obtain features from both views.
The temporal graph-convolution branch (T-GCN) and the frequency branch (F-GCN) are defined as follows:
\[
\tilde{E}_{i}=%
  \varepsilon\!\bigl(
      D_{i}^{-\tfrac12}A_{i}D_{i}^{-\tfrac12}
      \varepsilon(E_{i}W_{i1})W_{i2}+E_{i}
  \bigr)\!,\,%
  i\!\in\!\{T,F\}\kern0.4em\tag{4}
\]

Where $\tilde{x}_{i}\in\mathbb{R}^{C\times D_{i}}$ are hidden features of each view,
with $D_{i}$ denoting the degrees of $A_{i}$.
$W_{i1},W_{i2}\in\mathbb{R}^{D_{i}\times D_{i}'},\;$
are weight matrices, where $D_{i}'$ is adjustable hyper-parameters,
and $\varepsilon(\cdot)$ denote batch
normalization followed by ELU non-linear functions.
\begin{figure}
    \centering
    \includegraphics[width=1.0\linewidth]{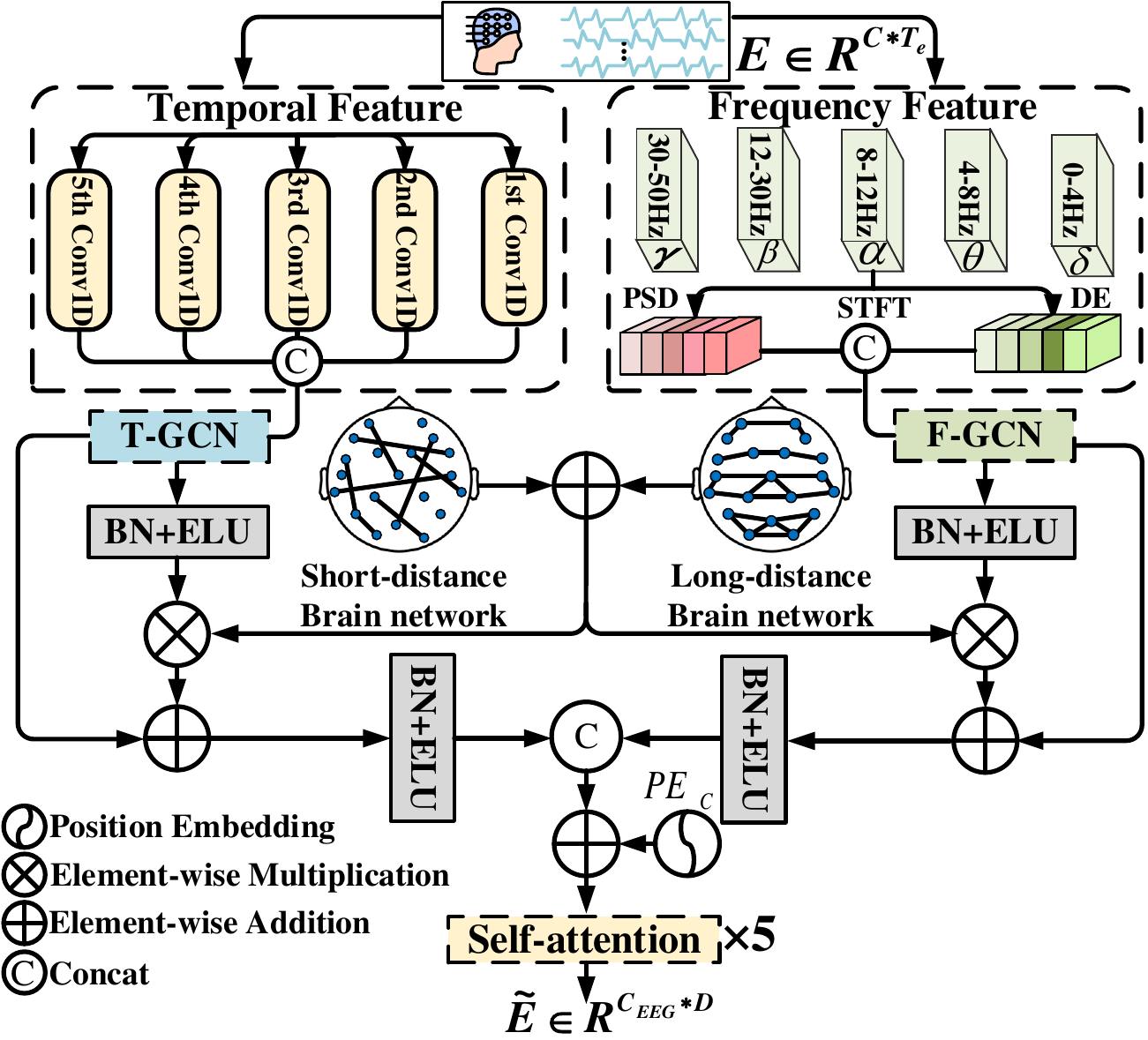}
    \caption{The overall block diagram of the proposed EEGEncode.}
    \label{fig:placeholder}
\end{figure}
Finally, the temporal and frequency features are concatenated along the feature dimension and then sent into a self-attention \cite{Vaswani2017Attention} mechanism to capture global features:
\begin{equation}
\widetilde{E}
  = \mathrm{SA}\!\left(
        \mathrm{PE}_{C}\!\left(
          \operatorname{Concat}\!\bigl(\widetilde{E}_{T},\widetilde{E}_{F}\bigr)
        \right)
    \right)
  \in \mathbb{R}^{B \times C_{EEG} \times D}
  \tag{5}
\end{equation}

\textbf{Speaker Extraction Network.}
The speaker extraction module is designed to estimate a mask $M$ that allows only the attended speaker’s voice to pass through $X'$, and the masked speech embedding $\hat{S}$ is obtained by:
\begin{equation}
  \hat{S} = X' \otimes M \in \mathbb{R}^{B \times C \times D}
  \tag{6}
\end{equation}

We concatenate the mixed audio features $X'$ and EEG features $\tilde{E}$ along the channel dimension, and then apply a 1D convolution to merge these features:
\begin{equation}
  X'_{\text{fuse}} = \operatorname{Conv1D}\!\bigl(\operatorname{concat}(X', \tilde{E})\bigr)
  \in \mathbb{R}^{B \times C \times D}
  \tag{7}
\end{equation}

The fused features are then fed into the speech separation models
to extract the mask $M$.

Inspired by speech separation models based on Temporal Convolutional Networks (TCN) \cite{Bai2018TCN} and Dual-Path RNNs (DPRNN) \cite{Luo2020DPRNN}, we adopt the more advanced MossFormer2 \cite{Zhao2024MossFormer2} as our separator. Unlike TCN and DPRNN, which are limited in long-term dependency modeling and computational efficiency, MossFormer2 combines local feature modeling with long-range contextual information. This balanced architecture enhances the extraction of target speech while suppressing irrelevant sources.

The MossFormer2 consists of two modules: the MossFormer 
Module and the RNN-Free Recurrent Module, which together model global 
context and local temporal patterns.

For the MossFormer module, it applies local full attention within non-overlapping 
chunks and linearized attention over the entire sequence. The fused output 
is refined by a gated convolutional unit:
\begin{equation}
  O = X'' + \operatorname{ConvM}\!\bigl(
        \sigma\!\bigl((U \otimes AV)\! \otimes\! AU\bigr)
      \bigr)
  \tag{8}
\end{equation}
where $X''$ is the block input, $U,V$ are projected features, $A$ is the attention map, $\sigma(\cdot)$ denotes the element-wise 
sigmoid gating, $\operatorname{ConvM}(\cdot)$ is a convolutional gating 
module, and $\otimes$ indicates element-wise multiplication.

For the RNN-free recurrent module, it employs a dilated feed-forward sequential 
memory network (FSMN) with gated convolutional units to model temporal 
recurrence:
\begin{equation}
  U = \operatorname{ConvU}(X), \quad V = \operatorname{ConvU}(X)
  \tag{9}
\end{equation}
\begin{equation}
  Y = \operatorname{DilatedFSMN}(V), \quad O = X + (U \otimes Y)
  \tag{10}
\end{equation}
Here, $X$ is the projected input to the recurrent block; 
$\operatorname{ConvU}(\cdot)$ denotes a point-wise convolutional unit 
used to form gates; DilatedFSMN() is a dilated feed-forward memory block with dense connections that aggregates contexts. This parallel design 
enlarges the receptive field and avoids sequential dependencies, enabling 
efficient real-time separation.

\textbf{Speech Decoder.}
The separated feature sequence is finally decoded into a waveform by the decoder:
\[\hat{s} = \mathrm{deconv1D}(\hat{S}) \in \mathbb{R}^{B \times 1 \times T_{s}}
\tag{11}
\]
The decoder is a 1D transposed convolutional layer, and it uses the
same kernel size and stride as the encoder.

\subsection{Loss Function}
\label{ssec:Loss Function}
In this study, we adopt the negative SI-SDR as the loss function, owing to its consistently good performance and its widespread use in target-speaker extraction. The SI-SDR is defined as:
\begin{equation}
\text{SI-SDR}
  = 10 \log_{10}
    \frac{\displaystyle
          \bigl\lVert
            \frac{\hat{s}^{\top}s}{\lVert s\rVert^{2}}\,s
          \bigr\rVert^{2}}
         {\displaystyle
          \bigl\lVert
            \frac{\hat{s}^{\top}s}{\lVert s\rVert^{2}}\,s
            - \hat{s}
          \bigr\rVert^{2}}
 \tag{12}
\end{equation}

where $\hat{s}$ and $s$ denote the extracted and true target-speaker signals, respectively.

\section{EXPERIMENTS}
\label{sec:pagestyle}

\subsection{Datasets}
\label{ssec:Datasets}

\textbf{\hspace*{\parindent}Cocktail Party Dataset.}
The first dataset \cite{Broderick2018SemanticEEG} used in this experiment comprises 33 adults (28 male, 5 female; 27.3 ± 3.2 years) with normal hearing and no neurological disorders. Each subject undergoes 30 trials, each lasting 60 seconds, where they listen to two different stories--one in each ear--narrated by different male speakers. Subjects are divided into two groups: one focusing on the left ear (17 subjects) and the other on the right ear (16 subjects, with one subject excluded).

\textbf{KUL Dataset.} The KUL dataset \cite{Biesmans2017} comprises 16 normal-hearing
participants, each completing 20 dichotic trials. We use the first
8 trials in which each subject participated, in which subjects were presented with different speech in the left and right ears. Participants are asked to pay attention to the sounds in one ear and ignore the sounds in the other. The BioSemi ActiveTwo system is used to record 64-channel EEG signals at an 8196 Hz sample rate.
\subsection{Data Processing}
\label{ssec:Data Processing}
 For the Cocktail Party data, our preprocessing steps remain consistent with UBESD: band-pass filtering (0.1--45 Hz), spline repair of noisy channels, mastoid-average rereferencing. For the KUL dataset, EEG data are first notch-filtered at 50~Hz to suppress power-line noise, then band-pass filtered (0.1--45~Hz, fourth-order Butterworth), re-referenced to the common average.
For both datasets, the EEG data are downsampled to 128Hz and cleaned with ICA to remove ocular and muscular artefacts, while the speech sampling rate is set to 44.1kHz.

\subsection{Implementation Details}
\label{ssec:Implementation Details}
For the Cocktail Party dataset, five trials are randomly selected as the test set, two trials are used for validation, and the remaining trials are used for training for each subject. For the KUL dataset, each subject’s trials are divided into training, validation, and test sets with proportions of $75\%$, $12.5\%$, and $12.5\%$, respectively.

Experiments were conducted using the \textsc{PyTorch} framework on an NVIDIA GeForce~4090 GPU. All models were trained for 60~epochs with a batch size of~1. The Adam optimizer was employed with a maximum learning rate of 0.0001.
A StepLR scheduler reduced the learning rate by a factor of 0.5 every 20~epochs, producing a piecewise-constant decay throughout training.

For the model implementation, the kernel sizes are set to $20$ for the Speech Encoder and Decoder. The encoder output dimension $C$ is set to $128$, and the number of separator blocks $R$ is set to $6$.

\subsection{Evaluation Metrics}
\label{ssec:Evaluation Metrics}

We assess our method with four metrics:SI-SDR(dB),PESQ \cite{Rix2001PESQ}, STOI \cite{Taal2010STOI}, and ESTOI \cite{Jensen2016ESTOI}. SI-SDR quantifies reconstruction fidelity; PESQ measures perceptual speech quality; STOI evaluates intelligibility via time--frequency correlations; ESTOI refines STOI for noisy conditions. Higher scores on all metrics indicate superior performance.

\section{RESULTS}
\label{sec:typestyle}

\subsection{Comparative Analysis}
\label{ssec:Comparative Analysis}
To validate the effectiveness of the proposed algorithm, we perform experiments on the Cocktail Party and the KUL dataset. First, we compare our method with baseline models. Then we analyze SI-SDR improvement variations across different EEG Encoder and speaker extraction networks.
\begin{table}[htbp]
  \centering
  \footnotesize
  \setlength{\tabcolsep}{3.5pt}

  \caption{Performance comparison on Cocktail~Party and KUL datasets}
  \label{tab:baseline}

  \begin{tabular}{c c c S[table-format=1.2] c S[table-format=1.2]}
    \toprule
    \textbf{Dataset} & \textbf{Model} & \textbf{SI--SDR\,(dB)} &
    {\textbf{STOI}} & {\textbf{ESTOI}} & {\textbf{PESQ}} \\
    \midrule
    \multirow{6}{*}{\makecell[l]{Cocktail~Party\\dataset}}
        & Mixture   & 0.45  & 0.71 & 0.55 & 1.61 \\
        & UBESD \cite{Hosseini2022BrainDriven}     & 8.54  & 0.83 & --   & 1.97 \\
        & BASEN \cite{Zhang2023BASEN}     & 11.56 & 0.86 & 0.72 & 2.21 \\
        & M3ANet \cite{Fan2025M3ANet}    & 13.95 & 0.89 & 0.78 & 2.58 \\
        & TFGA-Net(Ours)     & 15.91 & 0.92 & 0.82 & 2.36 \\
    \midrule
    \multirow{5}{*}{KUL}
        & Mixture   & 0.25 & 0.69 & 0.52 & 1.17 \\
        & UBESD \cite{Hosseini2022BrainDriven}    & 6.1  & 0.73 & 0.75 & 1.09 \\
        & BASEN \cite{Zhang2023BASEN}    & 11.5 & 0.82 & 0.76 & 1.76 \\
        & NeuroHeed \cite{Pan2024NeuroHeed} & 14.6 & 0.83 & 0.76 & 2.12 \\
        & TFGA-Net(Ours)     & 16.9 & 0.87 & 0.78 & 2.17 \\
    \bottomrule
  \end{tabular}
\end{table}

As shown in Table~I, the proposed TFGA-Net model attains state-of-the-art performance on the Cocktail~Party dataset, achieving \(15.91\,\text{dB}\) SI--SDR. Relative to UBESD, BASEN, and M3ANet, SI--SDR improves by \(7.37\), \(4.35\), \(3.02\), and \(1.96\,\text{dB}\), respectively. Compared with the state-of-the-art M3ANet method, our model achieves relative improvements of \(0.03\) and \(0.04\) in STOI and ESTOI, respectively, while reaching a modest PESQ score of \(2.36\). Against NeuroHeed on the KUL dataset, the TFGA-Net model delivers relative improvements of \(2.3\,\text{dB}\), \(0.04\), \(0.02\), and \(0.05\) in SI--SDR, STOI, ESTOI, and PESQ, respectively. Therefore, the TFGA-Net model offers competitive performance compared with existing brain-controlled speaker-extraction approaches.

\subsection{Ablation Study}
\label{ssec:Ablation Study}
To validate the contribution of each key module in TFGA-Net, we conduct ablation experiments on the Cocktail Party dataset. 
\begin{table}[htbp]
  \centering
  \footnotesize                       
  \setlength{\tabcolsep}{1.5pt}         
  \caption{Ablation experiments on the Cocktail~Party dataset}
  \label{tab:mossformer}

  \begin{tabular}{c c
                  c
                  S[table-format=1.2]
                  S[table-format=1.2]
                  S[table-format=1.2]}
    \toprule
    \textbf{Model} & \textbf{EEG Encoder} &
    {\textbf{SI--SDR}\,(dB)} &
    {\textbf{STOI}} &
    {\textbf{ESTOI}} &
    {\textbf{PESQ}} \\ \midrule
    Mixture      & -   & 0.45 & 0.71 & 0.55 & 1.61 \\
    TFGA-Net(Envelope)      & Envelope   & 10.24 & 0.78 & 0.69 & 1.66 \\
    TFGA-Net(T\textendash GCN) & T\textendash GCN & 14.78 & 0.86 & 0.73 & 1.91 \\
    TFGA-Net(F\textendash GCN) & F\textendash GCN & 14.72 & 0.86 & 0.72 & 1.90 \\
    TFGA-Net(ours) & TF\textendash GCN & 15.91 & 0.92 & 0.82 & 2.36 \\ \bottomrule
  \end{tabular}
\end{table}

\textbf{Ablation Study on EEGEncoder.}
To validate the performance of the temporal-frequency graph attention EEG-encoding framework, we conducted a controlled experiment in which only the EEG encoder varied. Specifically, the Envelope model feeds raw EEG signals directly without feature extraction; the T-GCN and F--GCN models extract temporal and spectral features, respectively. The experimental results are summarized in Table II. Relative to the Envelope, T--GCN, and F--GCN based encoders, the TF-GCN model yields SI--SDR gains of \(5.67\,\text{dB}\), \(1.13\,\text{dB}\), and \(1.19\,\text{dB}\), respectively, confirming its superiority. 

\textbf{Ablation Study on Speaker Extraction Network.}
\begin{figure}[htbp]
  \centering
  \begin{subfigure}[t]{0.495\linewidth}
    \centering
    \includegraphics[width=\linewidth]{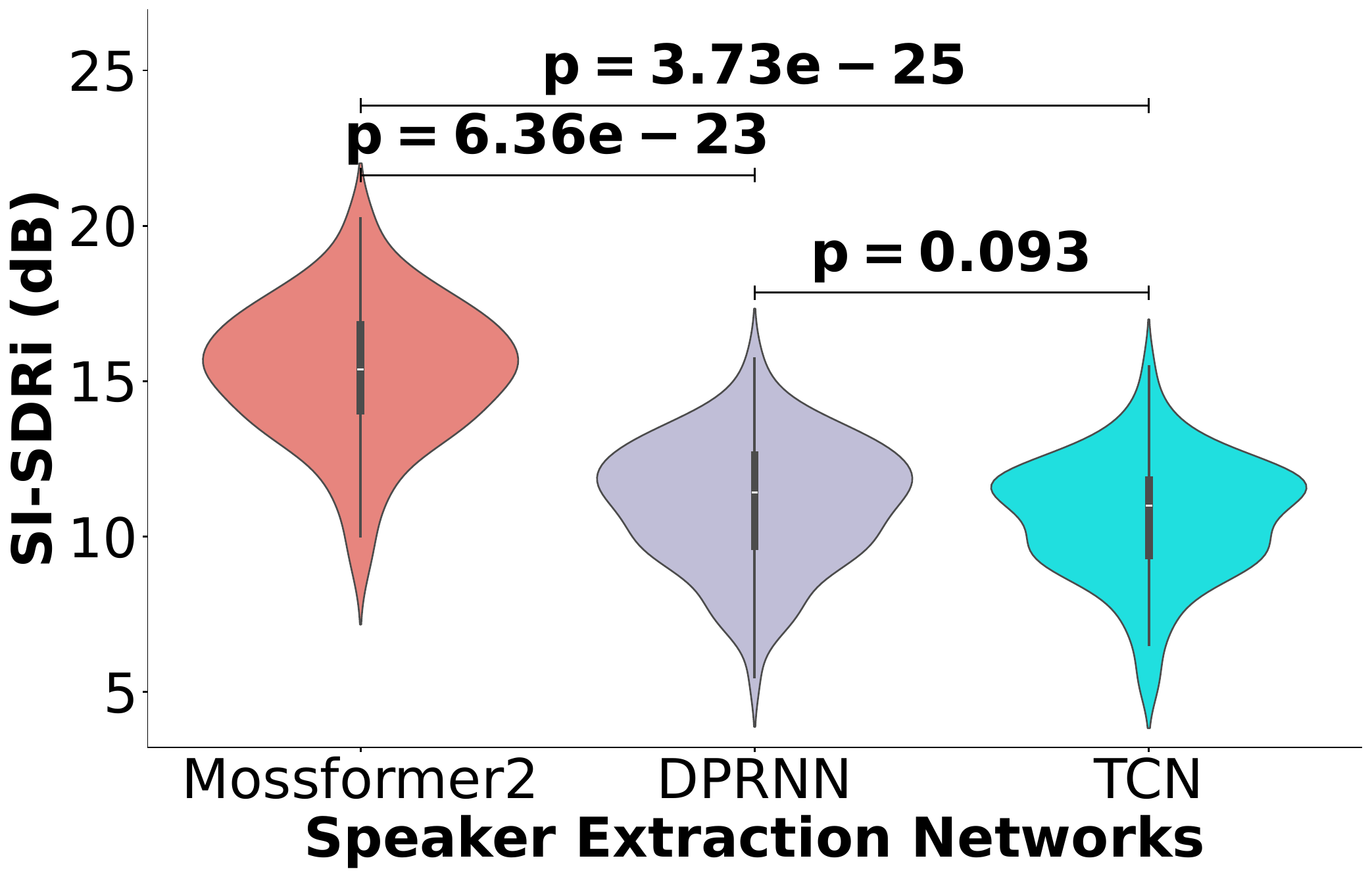}
    \vspace{-6pt}                
    \caption{Violin plot of SI-SDRi results}
    \label{fig:sisdri}
  \end{subfigure}%
  \hspace{0.01\linewidth}
  \begin{subfigure}[t]{0.495\linewidth}
    \centering
    \includegraphics[width=\linewidth]{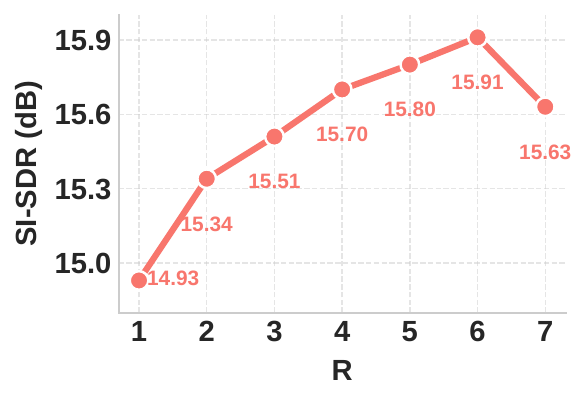}
    \vspace{-8pt}                
    \caption{MossFormer2 layers}
    \label{fig:layer}
  \end{subfigure}
  \caption{Performance on Speaker Extraction Network}
\end{figure}
 
 Fig.3(a) presents the violin plot of SI-SDRi distributions for the different speaker separation modules.The results demonstrate that the TFGA-Net model achieves superior overall performance.To fine-tune the optimal number of Mossformer2 layers, we evaluated the impact of varying the depth from~1 to~7. The results are summarized in Fig.3(b) Overall, increasing the number of layers significantly improves model performance, and the 6-layer MossFormer2 achieves the best results.

\section{CONCLUSION}
\label{sec:CONCLUSION}
In this paper, we propose a network that efficiently extracts multi-scale time--frequency features and incorporates cortical topological structures selectively engaged during the task. During the post-fusion feature processing stage , the network preserves the global context of the fused representations and, under EEG guidance, captures speech periodicity and prosody. In two-speaker scenarios, the results suggest that TFGA-Net achieves higher signal fidelity and better perceptual quality than current state-of-the-art methods. Experiments show that temporal-frequency graph attention network with MossFormer2 for EEG extraction outperforms T--GCN and F--GCN approaches. The MossFormer2 module further improves local--global dependency modeling under EEG guidance, achieving higher SI--SDR scores and lower variance than TCN and DPRNN.

\vfill\pagebreak




\bibliographystyle{IEEEbib}
\bibliography{strings,refs}

\end{document}